\documentstyle[seceq,epsf]{ptptex}

\def\sst{\scriptscriptstyle}

\title{
Three-Body-Cluster Effects on ${\it \Lambda}$ Single-Particle Energies\\
in $_{\it \Lambda}^{17}$O and $_{\it \Lambda}^{41}$Ca
}

\author{
Shinichiro {\sc Fujii},\footnote{
E-mail address: fuji2scp@mbox.nc.kyushu-u.ac.jp}
Ryoji {\sc Okamoto}$^{*,}$\footnote{
E-mail address: okamoto@mns.kyutech.ac.jp}
and Kenji {\sc Suzuki}$^{*,}$\footnote{
E-mail address: suzuki@mns.kyutech.ac.jp}
}

\inst{
Department of Physics, Kyushu University,
Fukuoka 812-8581, Japan
\\
$^{*}$Department of Physics, Kyushu Institute of Technology\\
Kitakyushu 804-8550, Japan
}

\abst{

A method for a microscopic description of ${\it \Lambda}$ hypernuclei
is formulated in the framework of the unitary-model-operator approach.
A unitarily transformed hamiltonian is introduced and given
in a cluster expansion form.
The structure of three-body-cluster terms are discussed especially
on the ${\Lambda}$ single-particle energy.
The ${\it \Lambda}$ single-particle energies including
the three-body-cluster contributions are calculated for the
$0s_{1/2}$, $0p_{3/2}$ and $0p_{1/2}$ states in $_{\it \Lambda}^{17}$O,
and for the $0s_{1/2}$, $0p_{3/2}$, $0p_{1/2}$, $0d_{5/2}$, $0d_{3/2}$
and $1s_{1/2}$ states in $_{\it \Lambda}^{41}$Ca,
using the Nijmegen soft-core (NSC), NSC97a-f, the J\"ulich \~A (J\~A)
and J\~B hyperon-nucleon interactions.
It is indicated that the three-body-cluster terms bring about
sizable effects in the magnitudes of the ${\it \Lambda}$
single-particle energies,
but hardly affect the ${\it \Lambda}$ spin-orbit splittings.
}

\begin{document}

\maketitle

\setcounter{equation}{0}

\section{Introduction}

Much attention has been paid to the study of hypernuclear physics
in recent years.
One of the fundamental problems in hypernuclear physics is to
understand properties of hypernuclei starting from the basic
interaction between two kinds of constituent particles,
nucleon (${\it N}$) and hyperon (${\it Y}$). 
Our knowledge 
on the $ {\it YN} $ interaction is, however, quite inadequate
because of the very limited experimental information on the ${\it YN}$
scattering.
In such a situation many-body theoretical studies of the hypernuclear
structure could provide alternative information on the ${\it YN}$
interaction. 
For this purpose, it would be desirable to have a theoretical method
which enables us to make a high-precision calculation.

In many of microscopic structure calculations of hypernuclei,
the $G$-matrix has been introduced as a ${\it YN}$ effective
interaction.
The $G$-matrix approach has been widely applied, and the use of this
type of effective interaction has made important contributions to the
description of
hypernuclei.\cite{rf:BAN90}\tocite{rf:VID98}
However, we may say from a formal point of view that the $G$-matrix is
energy-dependent and non-hermitian, and does not have
the property of decoupling between 
a model space of low-lying two-particle states and its complement.
The $G$-matrix itself is not considered to be an effective 
interaction in a formal sense of the effective interaction theory
because it does not have the property of decoupling.\cite{rf:KL} \ 
In order to derive such an effective
interaction we should add some higher-order corrections
such as folded diagrams.\cite{rf:KO,rf:TTKL99} \ 
It would be desirable if we could have a theory to describe
many-body systems in terms of an energy-independent and hermitian
effective interaction with the property of decoupling.

We have been proposed a many-body theory, the unitary-model-operator
approach (UMOA),\cite{rf:SO} \ 
that was formulated on the basis of such an effective interaction.
The UMOA was applied to finite nuclei $^{16}$O\cite{rf:SO}
and $^{40}$Ca,\cite{rf:KSO}
and some encouraging results were obtained.
Furthermore, in the previous work,\cite{rf:FOS99} \ 
we extended the formulation of the UMOA
to a description of ${\it \Lambda}$ hypernuclei and made calculations
of properties of $_{\it \Lambda}^{17}$O which is a typical closed-shell
nucleus plus one ${\Lambda}$ system,
using some realistic ${\it YN}$ interactions.

The formulation of the UMOA is based essentially on a cluster expansion
of a unitarily transformed hamiltonian in terms of a two-body
correlation operator.\cite{rf:PS,rf:SO86}
In the previous work we gave a method of treating correlations
within the two-body-cluster term, and 
the three-or-more-body-cluster terms remain to be evaluated.
It has been known that the expansion of the unitarily transformed
hamiltonian does not, in principle, terminate in finite order.
Therefore, it needs to give a method of evaluating the
three-or-more-body-cluster terms and confirm numerically convergence of
the cluster expansion.
It is of high interest to investigate to what extent
the three-body-cluster term would affect the ${\it \Lambda}$
single-particle energy, especially,
the ${\it \Lambda}$ spin-orbit splitting.
The calculation of the ${\it \Lambda}$ spin-orbit splitting is one of
the important tasks in hypernuclear physics because it has
a possibility of providing information to the determination of
some parameters in the ${\it YN}$ interaction such as the strengths of
the spin-orbit and antisymmetric spin-orbit components.

The purposes of this work are to formulate a method of 
evaluating the three-body-cluster effect in the framework of the UMOA,
and apply it to $^{17}_{\it \Lambda }$O
and $^{41}_{\it \Lambda }$Ca by employing various realistic ${\it YN}$
interactions.
This paper is organized as follows: In \S 2 we present a 
formulation of the UMOA to treat the three-body-cluster term. 
In \S 3 we apply the UMOA to $^{17}_{\it \Lambda }$O
and $^{41}_{\it \Lambda }$Ca by employing the Nijmegen
soft-core (NSC),\cite{rf:NSC} \ NSC97a-f,\cite{rf:NSC97} \ 
the J\"ulich \~A (J\~A) and J\~B\cite{rf:JUL} ${\it YN}$ potentials.
In \S 4 we make some concluding remarks.

\section{Formulation}

\subsection{Cluster expansion of the unitarily transformed hamiltonian}
\setcounter{equation}{0}

We first consider a hamiltonian of a hypernuclear system consisting of
nucleons and one ${\it \Lambda}$ (or ${\it \Sigma}$)
in the second-quantization form as

\begin{eqnarray}
\label{eq:1}
H&=&\sum_{\alpha \beta }
\langle \alpha |t_{\it N_{\rm 1}}|\beta \rangle c_{\alpha }^{\sst \dag }
c_{\beta }
+\frac{1}{4}\sum_{\alpha \beta \gamma \delta }
\langle \alpha \beta |v_{\it N_{\rm 1}N_{\rm 2}}|\gamma \delta \rangle
c_{\alpha }^{\sst \dag }c_{\beta }^{\sst \dag }c_{\delta }
c_{\gamma }\nonumber \\
&&+\sum_{\stackrel {\scriptstyle \mu \nu }{\it Y=\Lambda,\Sigma}}
\langle \mu |t_{\it Y}+\Delta m_{\it Y}|\nu \rangle 
d_{\mu }^{\sst \dag }d_{\nu }
+\sum_{\stackrel {\scriptstyle \mu \alpha \nu \beta }
{\it Y=\Lambda ,\Sigma}}
\langle \mu \alpha |v_{\it YN_{\rm 1}}|\nu \beta \rangle
d_{\mu }^{\sst \dag }c_{\alpha }^{\sst \dag }c_{\beta }d_{\nu },
\end{eqnarray}
where $c^{\sst \dag }$ ($c$) is the creation (annihilation)
operator for a nucleon
in the usual notation, and $d^{\sst \dag }$ $(d)$ is the creation
(annihilation)
operator for a hyperon, ${\it \Lambda}$ or ${\it \Sigma}$.
The kinetic energies of a nucleon and a hyperon are denoted by
$t_{\it N_{\rm 1}}$ and $t_{\it Y}$, respectively.
The $v_{\it N_{\rm 1}N_{\rm 2}}$ and
$v_{\it YN_{\rm 1}}$
represent the nucleon-nucleon (${\it NN}$) and
hyperon-nucleon (${\it YN}$) interactions, respectively.
The notations $\alpha ,\beta, \gamma$ and $\delta$ are used for
the sets of quantum numbers of nucleon states,
and $\mu$ and $\nu$ for those of hyperon states.
We note here that
$|\alpha \beta \rangle $ is
the antisymmetrized and normalized two-body ${\it NN}$ state.
The $\Delta m_{\it Y}$ denotes the difference between
the rest masses of ${\it \Lambda}$ and ${\it \Sigma}$.

In nuclear many-body problems, it is important to treat properly
the short-range two-body correlation because realistic
${\it NN}$ and ${\it YN}$ interactions have strongly repulsive cores
at short distance.
For this purpose, we introduce a unitary transformation of
the hamiltonian as
\begin{eqnarray}
\label{eq:2}
\bar{H}=e^{-S}He^{S},
\end{eqnarray}
where $S$ is the sum of anti-hermitian two-body operators for
${\it NN}$ and ${\it YN}$ systems defined as 
\begin{eqnarray}
\label{eq:3}
S=S^{(\it NN)}+S^{({\it YN})}
\end{eqnarray}
with
\begin{eqnarray}
\label{eq:4}
S^{(\it NN)}=\frac{1}{4}\sum_{\alpha \beta \gamma \delta}
\langle \alpha \beta |S_{\it N_{\rm 1}N_{\rm 2}}|\gamma \delta \rangle
c_{\alpha }^{\sst \dag }c_{\beta }^{\sst \dag }c_{\delta }c_{\gamma },
\end{eqnarray}
\begin{eqnarray}
\label{eq:5}
S^{(\it YN)}=\sum_{\mu \alpha \nu \beta}
\langle \mu \alpha |S_{\it YN_{\rm 1}}|\nu \beta \rangle
d_{\mu }^{\sst \dag }c_{\alpha }^{\sst \dag }c_{\beta }d_{\nu }.
\end{eqnarray}

According to Provid\^encia and Shakin,\cite{rf:PS} \ 
we make cluster expansion of the unitarily transformed hamiltonian as

\begin{eqnarray}
\label{eq:6}
\bar{H}=\bar{H}^{(1)}+\bar{H}^{(2)}+\bar{H}^{(3)}+\cdot \cdot \cdot ,
\end{eqnarray}
where the first three terms are written explicitly as
\begin{eqnarray}
\label{eq:7}
\bar{H}^{(1)}=\sum_{\alpha \beta }
\langle \alpha |h_{\it N_{\rm 1}}|\beta \rangle 
c_{\alpha }^{\sst \dag }c_{\beta }
+\sum_{\stackrel {\scriptstyle \mu \nu }{\it Y=\Lambda,\Sigma}}
\langle \mu |h_{\it Y}|\nu \rangle d_{\mu }^{\sst \dag }d_{\nu },
\end{eqnarray}
\begin{eqnarray}
\label{eq:8}
\bar{H}^{(2)}&=&\left( \frac{1}{2!}\right) ^{2}
\sum_{\alpha \beta \gamma \delta}
\langle \alpha \beta |\tilde{v}_{\it N_{\rm 1}N_{\rm 2}}|
\gamma \delta \rangle
c_{\alpha }^{\sst \dag }c_{\beta }^{\sst \dag }c_{\delta }c_{\gamma }
-\sum_{\alpha \beta }
\langle \alpha |u_{\it N_{\rm 1}}|\beta \rangle 
c_{\alpha }^{\sst \dag }c_{\beta }\nonumber \\
&&+\sum_{\stackrel {\scriptstyle \mu \alpha \nu \beta }
{\it Y=\Lambda ,\Sigma}}
\langle \mu \alpha |\tilde{v}_{\it YN_{\rm 1}}|\nu \beta \rangle
d_{\mu }^{\sst \dag }c_{\alpha }^{\sst \dag }c_{\beta }d_{\nu }
-\sum_{\stackrel {\scriptstyle \mu \nu }{\it Y=\Lambda,\Sigma}}
\langle \mu |u_{\it Y}|\nu \rangle d_{\mu }^{\sst \dag }d_{\nu }
\end{eqnarray}
and
\begin{eqnarray}
\label{eq:9}
\bar{H}^{(3)}&=&\left( \frac{1}{3!}\right) ^{2}
\sum_{\alpha \beta \gamma \delta \varepsilon \zeta}
\langle \alpha \beta \gamma |
\tilde{v}_{\it N_{\rm 1}N_{\rm 2}N_{\rm 3}}|
\delta \varepsilon \zeta \rangle
c_{\alpha }^{\sst \dag }c_{\beta }^{\sst \dag }c_{\gamma }^{\sst \dag }
c_{\zeta }c_{\varepsilon }c_{\delta }\nonumber \\
&&-\left( \frac{1}{2!}\right) ^{2}\sum_{\alpha \beta \gamma \delta}
\langle \alpha \beta |\tilde{u}_{\it N_{\rm 1}N_{\rm 2}}|
\gamma \delta \rangle
c_{\alpha }^{\sst \dag }c_{\beta }^{\sst \dag }c_{\delta }
c_{\gamma }\nonumber \\
&&+\left( \frac{1}{2!}\right) ^{2}
\sum_{\stackrel {\scriptstyle \mu \alpha \beta \nu \gamma \delta }
{\it Y=\Lambda ,\Sigma }}
\langle \mu \alpha \beta |\tilde{v}_{\it YN_{\rm 1}N_{\rm 2}}|
\nu \gamma \delta \rangle
d_{\mu }^{\sst \dag }c_{\alpha }^{\sst \dag }c_{\beta }^{\sst \dag }
c_{\delta }c_{\gamma }d_{\nu }\nonumber \\
&&-\sum_{\stackrel {\scriptstyle \mu \alpha \nu \beta }
{\it Y=\Lambda ,\Sigma}}
\langle \mu \alpha |\tilde{u}_{\it YN_{\rm 1}}|\nu \beta \rangle
d_{\mu }^{\sst \dag }c_{\alpha }^{\sst \dag }c_{\beta }d_{\nu }.
\end{eqnarray}
Here $|\alpha \beta \gamma \rangle$ is the antisymmetrized and
normalized three-body ${\it NNN}$ state,
and the ${\it NN}$ part of $|\mu \alpha \beta \rangle$ is
antisymmetrized and normalized.
In the above, we have used the notations given as follows:
Since the exponent $S$ is a two-body operator, the one-body operators
$h_{k}$ for $k={\it N_{\rm 1}}$ and ${\it Y}$ in $\bar{H}$ are
unchanged and given by
\begin{eqnarray}
\label{eq:10}
h_{\it N_{\rm 1}}=t_{\it N_{\rm 1}}+u_{\it N_{\rm 1}},
\end{eqnarray}
\begin{eqnarray}
\label{eq:11}
h_{\it Y}=t_{\it Y}+u_{\it Y}+\Delta m_{\it Y}.
\end{eqnarray}
The terms $u_{k}$ for $k={\it N_{\rm 1}}$ and ${\it Y}$ are
the auxiliary single-particle potentials of a nucleon and a hyperon,
respectively, and in this stage they are arbitrary.
The $\tilde{v}_{kl}$ for $\{kl\}=\{ {\it N_{\rm 1}N_{\rm 2}}\} $ and
$\{ {\it YN_{\rm 1}}\}$ are the transformed two-body interactions for
${\it NN}$ and ${\it YN}$ systems, and they are given by
\begin{eqnarray}
\label{eq:12}
\tilde{v}_{\it N_{\rm 1}N_{\rm 2}}=e^{-S_{\it N_{\rm 1}N_{\rm 2}}}
(h_{\it N_{\rm 1}}+h_{\it N_{\rm 2}}+v_{\it N_{\rm 1}N_{\rm 2}})
e^{S_{\it N_{\rm 1}N_{\rm 2}}}-(h_{\it N_{\rm 1}}+h_{\it N_{\rm 2}}),
\end{eqnarray}
\begin{eqnarray}
\label{eq:13}
\tilde{v}_{\it YN_{\rm 1}}=e^{-S_{\it YN_{\rm 1}}}
(h_{\it Y}+h_{\it N_{\rm 1}}+v_{\it YN_{\rm 1}})e^{S_{\it YN_{\rm 1}}}
-(h_{\it Y}+h_{\it N_{\rm 1}}).
\end{eqnarray}
The transformed hamiltonian $\bar{H}$ contains, in general,
three-or-more-body interactions even if the starting hamiltonian $H$
in Eq.~(\ref{eq:1}) does not include three-or-more-body interactions.
The transformed three-body interactions $\tilde{v}_{klm}$ for
$\{klm\}=\{ {\it N_{\rm 1}N_{\rm 2}N_{\rm 3}}\}$ and
$\{ {\it YN_{\rm 1}N_{\rm 2}}\}$ are given, respectively, by
\begin{eqnarray}
\label{eq:14}
\tilde{v}_{\it N_{\rm 1}N_{\rm 2}N_{\rm 3}}&=&
e^{-S_{\it N_{\rm 1}N_{\rm 2}N_{\rm 3}}}
(h_{\it N_{\rm 1}}+h_{\it N_{\rm 2}}+h_{\it N_{\rm 3}}
+v_{\it N_{\rm 1}N_{\rm 2}}+v_{\it N_{\rm 2}N_{\rm 3}}
+v_{\it N_{\rm 3}N_{\rm 1}})
e^{S_{\it N_{\rm 1}N_{\rm 2}N_{\rm 3}}} \nonumber \\
&&-(h_{\it N_{\rm 1}}+h_{\it N_{\rm 2}}+h_{\it N_{\rm 3}}
+\tilde{v}_{\it N_{\rm 1}N_{\rm 2}}+\tilde{v}_{\it N_{\rm 2}N_{\rm 3}}
+\tilde{v}_{\it N_{\rm 3}N_{\rm 1}}),
\end{eqnarray}
\begin{eqnarray}
\label{eq:15}
\tilde{v}_{\it YN_{\rm 1}N_{\rm 2}}&=&e^{-S_{\it YN_{\rm 1}N_{\rm 2}}}
(h_{\it Y}+h_{\it N_{\rm 1}}+h_{\it N_{\rm 2}}
+v_{\it YN_{\rm 1}}+v_{\it N_{\rm 1}N_{\rm 2}}+v_{\it N_{\rm 2}Y})
e^{S_{\it YN_{\rm 1}N_{\rm 2}}}\nonumber \\
&&-(h_{\it Y}+h_{\it N_{\rm 1}}+h_{\it N_{\rm 2}}
+\tilde{v}_{\it YN_{\rm 1}}+\tilde{v}_{\it N_{\rm 1}N_{\rm 2}}
+\tilde{v}_{\it N_{\rm 2}Y}),
\end{eqnarray}
where the operators $S_{klm}$ for
$\{klm\}=\{ {\it N_{\rm 1}N_{\rm 2}N_{\rm 3}}\}$ and
$\{ {\it YN_{\rm 1}N_{\rm 2}}\}$
denote the sum of the two-body correlation operators defined as
\begin{eqnarray}
\label{eq:18}
S_{\it N_{\rm 1}N_{\rm 2}N_{\rm 3}}=S_{\it N_{\rm 1}N_{\rm 2}}
+S_{\it N_{\rm 2}N_{\rm 3}}+S_{\it N_{\rm 3}N_{\rm 1}},
\end{eqnarray}
\begin{eqnarray}
\label{eq:19}
S_{\it YN_{\rm 1}N_{\rm 2}}=S_{\it YN_{\rm 1}}
+S_{\it N_{\rm 1}N_{\rm 2}}+S_{\it N_{\rm 2}Y}.
\end{eqnarray}
We here note that the unitary transformation of the auxiliary
potentials $u_{k}$ for
$k={\it N_{\rm 1}}$ (${\it N_{\rm 2}}$) and ${\it Y}$
generates two-body operators written as
\begin{eqnarray}
\label{eq:16}
\tilde{u}_{\it N_{\rm 1}N_{\rm 2}}=e^{-S_{\it N_{\rm 1}N_{\rm 2}}}
(u_{\it N_{\rm 1}}+u_{\it N_{\rm 2}})e^{S_{\it N_{\rm 1}N_{\rm 2}}}
-(u_{\it N_{\rm 1}}+u_{\it N_{\rm 2}}),
\end{eqnarray}
\begin{eqnarray}
\label{eq:17}
\tilde{u}_{\it YN_{\rm 1}}=e^{- S_{\it YN_{\rm 1}}}
(u_{\it Y}+u_{\it N_{\rm 1}})e^{S_{\it YN_{\rm 1}}}
-(u_{\it Y}+u_{\it N_{\rm 1}}).
\end{eqnarray}

In nuclear many-body problems, it is important how to choose
the auxiliary potential $u_{k}$.
In general, it is useful to introduce
$u_{\it N_{\rm 1}}$ ($u_{\it N_{\rm 2}}$) and $u_{\it Y}$ as
self-consistent potentials
defined with the transformed two-body interactions
$\tilde{v}_{\it N_{\rm 1}N_{\rm 2}}$ and
$\tilde{v}_{\it YN_{\rm 1}}$ as
\begin{eqnarray}
\label{eq:20}
\langle \alpha |u_{\it N_{\rm 1}}| \beta \rangle
=\sum_{\xi \leq \rho _{\rm F}}
\langle \alpha \xi |\tilde{v}_{\it N_{\rm 1}N_{\rm 2}}|
\beta \xi \rangle ,
\end{eqnarray}
\begin{eqnarray}
\label{eq:21}
\langle \mu |u_{\it Y}| \nu \rangle =\sum_{\xi \leq \rho _{\rm F}}
\langle \mu \xi |\tilde{v}_{\it YN_{\rm 1}}|\nu \xi \rangle,
\end{eqnarray}
where $\rho _{\rm F}$ is the uppermost occupied level, and the notation
$\xi$ means an occupied state for nucleons.
If $u_{\it N_{\rm 1}}$ and $u_{\it Y}$ are defined as given in
Eqs.~(\ref{eq:20}) and (\ref{eq:21}), one can prove
that the second and fourth terms on the right-hand side of
Eq.~(\ref{eq:8}) are cancelled by
the contributions of the bubble diagrams of the first and third terms,
respectively.
In the same way, the second and fourth terms on the right-hand side of
Eq.~(\ref{eq:9}) are also cancelled by the bubble-diagram contributions
of the first and third terms, respectively.
Therefore, the cluster terms $\bar{H}^{(1)}$, $\bar{H}^{(2)}$ and
$\bar{H}^{(3)}$ include only the one-, two- and three-body operators,
respectively,
if we write them in the normal-product form with respect to particles
and holes as discussed in Ref.~\citen{rf:SO}.
As has been well known, this cancellation mechanism is advantageous
in perturbative calculations, because there is no need of evaluating
${\it u}$-insertion diagrams and bubble diagrams.

An important problem in the present approach is how to determine the
two-body correlation operators
$S_{kl}$ for \{$kl$\} $=$ \{${\it N_{\rm 1}N_{\rm 2}}$\}
and \{${\it YN_{\rm 1}}$\}.
These operators are given as solutions to equations of decoupling as
\begin{eqnarray}
\label{eq:decoupling}
Q_{kl}e^{-S_{kl}}
(h_{k}+h_{l}
+\tilde{v}_{kl})e^{S_{kl}}P_{kl}=0,
\end{eqnarray}
where $P_{kl}$ and $Q_{kl}$ for \{$kl$\} $=$
\{${\it N_{\rm 1}N_{\rm 2}}$\} and \{${\it YN_{\rm 1}}$\}
are projection operators which act in the space of two-body states and
project a two-body state onto the low-momentum model space
and the high-momentum space, respectively.
It has been proved that Eq.~(\ref{eq:decoupling})
for $S_{kl}$ can be solved in a nonperturbative
way\cite{rf:SO,rf:FOS99} under the restrictive conditions
\begin{eqnarray}
\label{eq:restriction}
P_{kl}S_{kl}P_{kl}=Q_{kl}S_{kl}Q_{kl}=0.
\end{eqnarray}

In the previous work\cite{rf:FOS99} the structure of the
two-body-cluster term for the ${\it \Lambda}$ single-particle energy
was studied,
and three-or-more-body cluster terms were not considered.
After the evaluation of the two-body-cluster term, the next
important term will be the three-body-cluster (TBC) term.
The significance of studying the TBC term may be understood from the
following two reasons:
The first is to examine whether the TBC effect is large or not
in comparison with the one- and two-body-cluster effects,
in other words, whether the cluster expansion of $\bar{H}$ is
convergent so fast or not as to be useful in a practical application.
The second is that, when we investigate effects of the ${\it real}$
three-body force,
we must be careful of the possibility that such three-body-force effects may
interfere with effects of the ${\it effective}$ three-body force.

\subsection{Structure of the three-body-cluster term}

We here discuss general structure of the TBC term
$\bar{H}^{(3)}$ in Eq.~(\ref{eq:9}). The TBC term $\bar{H}^{(3)}$
contains the transformed
three-body ${\it NNN}$ and ${\it YNN}$ interactions.
These interactions include the bare two-body ${\it NN}$ and ${\it YN}$
interactions which have strongly repulsive cores.
Therefore, we represent the transformed three-body interactions
in terms of the transformed two-body ${\it NN}$ and ${\it YN}$
interactions $\tilde{v}_{\it N_{\rm 1}N_{\rm 2}}$ and
$\tilde{v}_{\it YN_{\rm 1}}$ which are well-behaved
in the actual calculation.

Using a factorization formula of Campbel and Hausdorf,
we have for exp($S_{ijk}$)
\begin{eqnarray}
\label{eq:22}
e^{S_{ijk}}=e^{S_{ij}}e^{S_{(ij)k}}
=e^{S_{jk}}e^{S_{(jk)i}}
=e^{S_{ki}}e^{S_{(ki)j}}
\end{eqnarray}
with
\begin{eqnarray}
\label{eq:23}
S_{(ij)k}&=&S_{jk}+S_{ki}-\frac{1}{2}
[S_{ij},S_{jk}+S_{ki}]\nonumber \\
&&-\frac{1}{12}[[S_{ij},S_{jk}+S_{ki}],S_{ijk}
+S_{ij}]+\cdot \cdot \cdot .
\end{eqnarray}
Hereafter, a set \{$ijk$\} denotes that of
\{${\it N_{\rm 1}N_{\rm 2}N_{\rm 3}}$\} or
\{${\it YN_{\rm 1}N_{\rm 2}}$\}.
The correlation operators $S_{(jk)i}$ and $S_{(ki)j}$ are also written
in a similar way.
Substituting exp($S_{ijk}$) in Eq.~(\ref{eq:22}) into
Eqs.~(\ref{eq:14}) and (\ref{eq:15}), we have
\begin{eqnarray}
\label{eq:24}
\tilde{v}_{ijk}=\tilde{v}_{ijk}^{(v)}+\tilde{v}_{ijk}^{(h)}
\end{eqnarray}
with
\begin{eqnarray}
\label{eq:25}
\tilde{v}_{ijk}^{(v)}=\sum_{(ijk)}(e^{-S_{(ij)k}}
\tilde{v}_{ij}e^{S_{(ij)k}}-\tilde{v}_{ij})
\end{eqnarray}
and
\begin{eqnarray}
\label{eq:26}
\tilde{v}_{ijk}^{(h)}&=&\sum_{(ijk)}[e^{-S_{(ij)k}}
(h_{i}+h_{j})
e^{S_{(ij)k}}-(h_{i}+h_{j})]\nonumber \\
&&+(h_{i}+h_{j}+h_{k})-e^{-S_{ijk}}(h_{i}+h_{j}+h_{k})
e^{S_{ijk}},
\end{eqnarray}
where $\sum_{(ijk)}$ denotes the summation of exchange terms
defined for an arbitrary function $f(ijk)$ as
\begin{eqnarray}
\label{eq:27}
\sum_{(ijk)}f(ijk)=f(ijk)+f(jki)+f(kij).
\end{eqnarray}

The terms $\tilde{v}_{ijk}^{(v)}$ and $\tilde{v}_{ijk}^{(h)}$
can be expanded into series in powers of $S_{ij}$, $S_{jk}$
and $S_{ki}$.
The term $\tilde{v}_{ijk}^{(v)}$ is given by
\begin{eqnarray}
\label{eq:28}
\tilde{v}_{ijk}^{(v)}&=&\sum_{(ijk)}
\{[\tilde{v}_{ij},S_{jk}+S_{ki}]-\frac{1}{2}
[\tilde{v}_{ij},[S_{ij},S_{jk}+S_{ki}]]\nonumber \\
&&+\frac{1}{2}[[\tilde{v}_{ij},S_{jk}+S_{ki}],S_{jk}+S_{ki}]\}
+\cdot \cdot \cdot .
\end{eqnarray}
Similarly, $\tilde{v}_{ijk}^{(h)}$ is given by
\begin{eqnarray}
\label{eq:29}
\tilde{v}_{ijk}^{(h)}&=&\sum_{(ijk)}\{-\frac{1}{2}
[[h_{i}+h_{j},S_{ij}],S_{jk}+S_{ki}]\nonumber \\
&&+\frac{1}{6}[[[h_{i}+h_{j},S_{ij}],S_{ij}],S_{jk}+S_{ki}]\nonumber \\
&&+\frac{1}{3}[[[h_{i}+h_{j},S_{ij}],S_{jk}+S_{ki}],S_{ijk}]\}
+\cdot \cdot \cdot .
\end{eqnarray}

The expressions of $\tilde{v}_{ijk}^{(v)}$ and $\tilde{v}_{ijk}^{(h)}$
give us desirable forms for the calculation of the TBC terms
because they do not include the bare interaction $\tilde{v}_{ij}$
and are written in terms of only
the terms $h_{i}$, $\tilde{v}_{ij}$ and $S_{ij}$.
The TBC effect on the ground-state and one-body energies for ordinary
nuclei such as $^{16}$O and $^{40}$Ca has already been
studied.\cite{rf:SO,rf:KSO,rf:SO86} \ 
In the present work we study the effect of the TBC term on the energy
of ${\it \Lambda}$.

\subsection{Derivation of the three-body-cluster contribution to the
${\it \Lambda}$ single-particle energy}

The contribution of the TBC term to a ${\it \Lambda}$ single-particle
energy in a state $\lambda _{\it \Lambda}$ is derived from
Eq.~(\ref{eq:9}) by making the summation
of diagonal occupied states of nucleons $\mu _{\it N}$ and
$\nu _{\it N}$ as
\begin{eqnarray}
\label{eq:30}
\Delta E_{\lambda _{\it \Lambda}}^{(\rm TBC)}&=&\frac{1}{2}
\sum_{\mu _{\it N}\nu _{\it N}\leq \rho _{\rm F}}
\langle \lambda _{\it \Lambda}\mu _{\it N}\nu _{\it N}|
\tilde{v}_{\it \Lambda N_{\rm 1}N_{\rm2}}
|\lambda _{\it \Lambda}\mu _{\it N}\nu _{\it N} \rangle \nonumber \\
&&-\sum_{\mu _{\it N}\rho _{\rm F}}
\langle \lambda _{\it \Lambda}\mu _{\it N}|
\tilde{u}_{\it \Lambda N_{\rm 1}}|
\lambda _{\it \Lambda}\mu _{\it N} \rangle ,\\ \nonumber
\end{eqnarray}
where $ \tilde{v}_{\it \Lambda N_{\rm 1}N_{\rm2}} $ is the transformed
three-body ${\it \Lambda NN}$ interaction $ \tilde{v}_{ijk} $ for
$\{ijk\}=\{{\it \Lambda N_{\rm 1}N_{\rm 2}}\}$ in Eq.~(\ref{eq:24}),
and $\tilde{u}_{\it \Lambda N_{\rm 1}}$ is the transformed operator
of the one-body potential defined for ${\it Y=\Lambda}$ in
Eq.~(\ref{eq:17}).

\begin{figure}[t]
  \label{fig:1}
  \epsfxsize = 10cm
  \centerline{\epsfbox{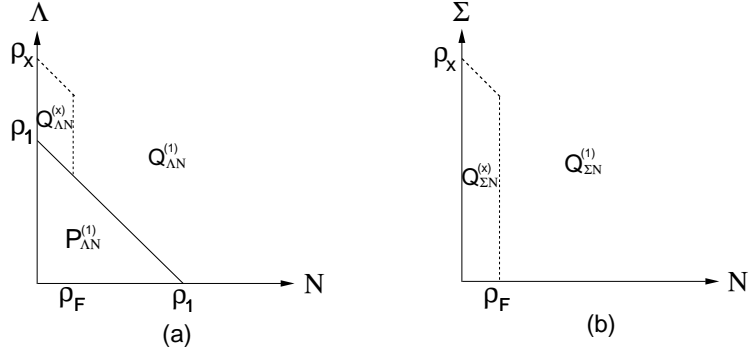}}
  \caption {The model space and its complement in the
 first-step calculation.}
\end{figure}

\begin{figure}[t]
  \epsfxsize = 5cm
  \centerline{\epsfbox{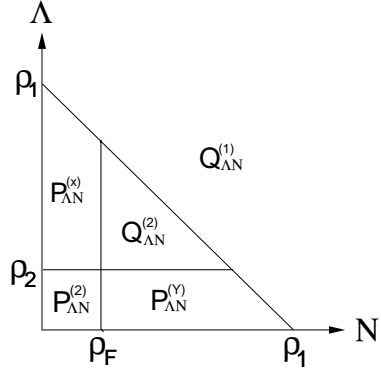}}
  \caption {The model space and its complement in the
 second-step calculation.}
  \label{fig:2}
\end{figure}


In order to obtain the TBC effect in Eq.~(\ref{eq:30}),
we need to calculate the matrix element of the two-body correlation
operator $S_{ij}$,
the two-body effective interaction $\tilde{v}_{ij}$ and
the single-particle potential $u_{k}$.
The calculation procedure for these quantities has already been given
in Refs.~\citen{rf:SO} and \citen{rf:FOS99} in detail.
Therefore, we here present only the outline of
the calculation procedure.
In an actual calculation for the ${\it YN}$ system,
we employ a two-step procedure by introducing 
two kinds of model spaces, namely, large and small model spaces
as shown in Figs. 1 and 2.
We here use the harmonic oscillator wave functions as basis states.
The numbers $\rho _{\rm F}$, $\rho _{\rm X}$, $\rho _{1}$ and
$\rho _{2}$
in Figs.~1 and 2 are defined as
\begin{eqnarray}
\rho _{\rm F}=2n_{\it N}+l_{\it N},
\end{eqnarray}
\begin{eqnarray}
\rho _{2}=2n_{\it \Lambda}+l_{\it \Lambda},
\end{eqnarray}
\begin{eqnarray}
\rho _{\rm X}=2n_{\it Y}+l_{\it Y}+2n_{\it N}+l_{\it N},
\ \ ({\it Y=\Lambda, \Sigma})
\end{eqnarray}
and
\begin{eqnarray}
\rho _{1}=2n_{\it \Lambda}+l_{\it \Lambda}+2n_{\it N}+l_{\it N},
\end{eqnarray}
where $n_{k}$ and $l_{k}$ for $k={\it \Lambda}$, ${\it \Sigma}$ and
${\it N}$ are
the harmonic oscillator quantum numbers.
The number $ \rho _{2} $ is introduced so as to specify
the uppermost bound state of $ {\it \Lambda } $.

In the first-step calculation, we calculate the ${\it \Lambda N}$
effective interaction $\tilde{v}_{\it \Lambda N}^{(1)}$ acting in
the large model space $P_{\it \Lambda N}^{(1)}$ by solving the
equation of decoupling between the $P_{\it \Lambda N}^{(1)}$ and the
$\{ (Q_{\it \Lambda N}^{(1)}-Q_{\it \Lambda N}^{(\rm X)})
+(Q_{\it \Sigma N}^{(1)}-Q_{\it \Sigma N}^{(\rm X)})\}$ spaces in
Eq.~(\ref{eq:decoupling}).\cite{rf:FOS99} \ 
It is noted here that the $Q_{\it \Lambda N}^{(\rm X)}$ and
$Q_{\it \Sigma N}^{(\rm X)}$
spaces are the excluded spaces due to the Pauli principle for nucleons.
The ${\it \Lambda}$ potential energy $u_{\it \Lambda}^{(1)}$
in the first step is calculated as given in Eq.~(\ref{eq:21}),
using the ${\it \Lambda N}$ effective interaction
$\tilde{v}_{\it \Lambda N}^{(1)}$.

\begin{figure}[t]
  \epsfxsize = 9cm
  \centerline{\epsfbox{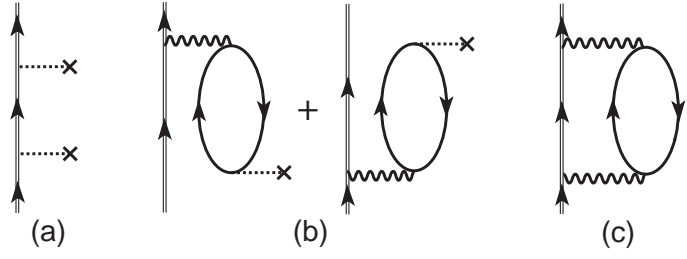}}
  \caption {Calculated diagrams in the second order for the
${\it \Lambda}$ single-particle energies.
 The $ {\it \Lambda N} $ effective interaction
$\tilde{v}_{{\it \Lambda N}}$ is
 represented by the wiggly-line vertex, and the vertex $\times $ with
the dashed line represents the non-diagonal part of the one-body
hamiltonian $ h_{k}=t_{k}+u_{k} $ for $ k={\it \Lambda } $ and
$ {\it N} $.
The propagations of $ {\it N} $ and $ {\it \Lambda } $ are represented
by the single and double external lines, respectively.}
  \label{fig:3}
\end{figure}

In the second-step calculation, the ${\it \Lambda N}$ effective
interaction $\tilde{v}_{\it \Lambda N}^{(2)}$ acting in the
$P_{\it \Lambda N}^{(2)}$ space and the correlation operator
$S_{\it \Lambda N}^{(2)}$ are obtained by solving the equation of
decoupling between the $P_{\it \Lambda N}^{(2)}$ and the
$Q_{\it \Lambda N}^{(2)}$ spaces,
using $\tilde{v}_{\it \Lambda N}^{(1)}$ and $u_{\it \Lambda}^{(1)}$.
The ${\it \Lambda}$ potential energy $u_{\it \Lambda}^{(2)}$ is
finally calculated in Eq.~(\ref{eq:21}),
using $\tilde{v}_{\it \Lambda N}^{(2)}$.
However, at this stage, we consider only the diagonal part of
$h_{\it \Lambda}^{(2)}=t_{\it \Lambda}+u_{\it \Lambda}^{(2)}$
as the unperturbed one-body hamiltonian.
Therefore, we evaluate perturbation corrections arising from the
non-diagonal terms of $h_{\it \Lambda}^{(2)}$ as shown in diagram (a)
of Fig.~3.
Furthermore, the effective interaction between ${\it \Lambda N}$
states in the $P_{\it \Lambda N}^{(2)}$ and
$P_{\it \Lambda N}^{({\rm Y})}$ spaces induces
core polarization in the closed-shell nucleus.
We take account of this effect by calculating
diagrams (b) and (c) of Fig.~3,
using the ${\it \Lambda N}$ effective interaction
$\tilde{v}_{\it \Lambda N}^{(1)}$ determined in the first step.
The validity of this two-step calculation procedure is confirmed
numerically in the previous work.\cite{rf:FOS99}

In order to calculate the TBC effect on the ${\it \Lambda}$
single-particle energy,
we need the effective interaction and the correlation operator
for the ${\it NN}$ system which are determined by solving the equation
of decoupling in Eq.~(\ref{eq:decoupling}).
Using the two-body effective interactions and the two-body correlation
operators for the ${\it \Lambda N}$ and ${\it NN}$ systems,
one can derive that an approximate expression of the TBC terms from
Eq.~(\ref{eq:30}) as
\begin{eqnarray}
\label{eq:34}
\Delta E_{\lambda _{\it \Lambda}}^{(\rm TBC)}&\simeq &\frac{1}{2}
\sum_{\stackrel {\scriptstyle \mu _{\it N}\nu _{\it N}
\leq \rho _{\rm F}}{({\it \Lambda N_{\rm 1}N_{\rm 2}})}}
\langle \lambda _{\it \Lambda}\mu _{\it N}\nu _{\it N}|
\frac{1}{2}[[\tilde{v}_{\it \Lambda N_{\rm 1}},
S_{\it N_{\rm 1}N_{\rm 2}}
+S_{\it N_{\rm 2}\Lambda }],S_{\it N_{\rm 1}N_{\rm 2}}
+S_{\it N_{\rm 2}\Lambda}]
|\lambda _{\it \Lambda}\mu _{\it N}\nu _{\it N} \rangle \nonumber \\
&=& (a_{\rm T})+(b_{\rm T})+(c_{\rm T})+(d_{\rm T})
+(e_{\rm T})+(f_{\rm T}),
\end{eqnarray}
where
\begin{eqnarray}
\label{eq:35}
(a_{\rm T})&=& -\sum_{\stackrel {\scriptstyle \mu _{\it N}\nu _{\it N}
\leq \rho _{\rm F}}{\nu _{\it \Lambda }\leq \rho_{2}}}
\sum_{\stackrel {\scriptstyle \alpha _{\it N} > \rho _{\rm F}}
{\alpha _{\it \Lambda } > \rho_{2}}}
\tilde{v}^{({\it \Lambda N})}_{\lambda _{\it \Lambda}\nu _{\it N}
\nu _{\it \Lambda }\mu _{\it N}}
S^{({\it \Lambda N})}_{\nu _{\it \Lambda}\mu _{\it N}
\alpha _{\it \Lambda }\alpha _{\it N}}
S^{({\it \Lambda N})}_{\alpha _{\it \Lambda}\alpha _{\it N}
\lambda _{\it \Lambda }\nu _{\it N}},\nonumber \\
(b_{\rm T})&=& \frac{1}{2}\sum_{\mu _{\it N}\nu _{\it N}
\rho _{\it N}\leq \rho _{\rm F}}
\sum_{\alpha _{\it N}\beta _{\it N} > \rho _{\rm F}}
\tilde{v}^{({\it \Lambda N})}_{\lambda _{\it \Lambda}\mu _{\it N}
\lambda _{\it \Lambda }\rho _{\it N}}
S^{({\it NN})}_{\rho _{\it N}\nu _{\it N}\alpha _{\it N}\beta _{\it N}}
S^{({\it NN})}_{\alpha _{\it N}
\beta _{\it N}\mu _{\it N}\nu _{\it N}},\nonumber \\
(c_{\rm T})&=& -2\sum_{\mu _{\it N}\nu _{\it N}\leq \rho _{\rm F}}
\sum_{\stackrel {\scriptstyle \alpha _{\it N}\beta _{\it N} >
\rho _{\rm F}}{\alpha _{\it \Lambda } > \rho_{2}}}
\tilde{v}^{({\it \Lambda N})}_{\lambda _{\it \Lambda}
\alpha _{\it N}\alpha _{\it \Lambda }\mu _{\it N}}
S^{({\it NN})}_{\mu _{\it N}\nu _{\it N}\alpha _{\it N}\beta _{\it N}}
S^{({\it \Lambda N})}_{\alpha _{\it \Lambda}\beta _{\it N}
\lambda _{\it \Lambda }\nu _{\it N}},\nonumber \\
(d_{\rm T})&=& \sum_{\mu _{\it N}\nu _{\it N}\leq \rho _{\rm F}}
\sum_{\stackrel {\scriptstyle \alpha _{\it N}\beta _{\it N} >
\rho _{\rm F}}{\alpha _{\it \Lambda}>\rho_{2}}}
\tilde{v}^{({\it NN})}_{\mu _{\it N}\alpha _{\it N}
\nu _{\it N}\beta _{\it N}}
S^{({\it \Lambda N})}_{\lambda_{\it \Lambda}
\nu _{\it N}\alpha _{\it \Lambda }\alpha _{\it N}}
S^{({\it \Lambda N})}_{\alpha_{\it \Lambda}\beta _{\it N}
\lambda _{\it \Lambda }\mu _{\it N}},\nonumber \\
(e_{\rm T})&=& \sum_{\mu _{\it N}\nu _{\it N}\leq \rho _{\rm F}}
\sum_{\stackrel {\scriptstyle \alpha _{\it N}> \rho _{\rm F}}
{\alpha _{\it \Lambda }\beta _{\it \Lambda } > \rho_{2}}}
\tilde{v}^{({\it \Lambda N})}_{\alpha _{\it \Lambda}\nu _{\it N}
\beta _{\it \Lambda }\mu _{\it N}}
S^{({\it \Lambda N})}_{\lambda _{\it \Lambda}\mu _{\it N}
\alpha _{\it \Lambda }\alpha _{\it N}}
S^{({\it \Lambda N})}_{\beta _{\it \Lambda}\alpha _{\it N}
\lambda _{\it \Lambda }\nu _{\it N}},\nonumber \\
(f_{\rm T})&=& -\frac{1}{2}\sum_{\mu _{\it N}\nu _{\it N}
\leq \rho _{\rm F}}
\sum_{\alpha _{\it N}\beta _{\it N}\gamma _{\it N}> \rho _{\rm F}}
\tilde{v}^{({\it \Lambda N})}_{\lambda _{\it \Lambda}
\alpha _{\it N}\lambda _{\it \Lambda }\gamma _{\it N}}
S^{({\it NN})}_{\mu _{\it N}\nu _{\it N}\alpha _{\it N}\beta _{\it N}}
S^{({\it NN})}_{\gamma _{\it N}\beta _{\it N}\mu _{\it N}\nu _{\it N}}.
\end{eqnarray}
In the above, we have used the notations defined by
\begin{eqnarray}
\label{eq:def_v}
\tilde{v}^{({\it \Lambda N})}_{\alpha _{\it \Lambda}\beta _{\it N}
\gamma _{\it \Lambda}\delta _{\it N}}
&=&\langle \alpha _{\it \Lambda}\beta _{\it N}|
\tilde{v}_{\it \Lambda N_{\rm 1}}|\gamma _{\it \Lambda }
\delta _{\it N}\rangle ,\nonumber \\
\tilde{v}^{({\it NN})}_{\alpha _{\it N}\beta _{\it N}\gamma _{\it N}
\delta _{\it N}}
&=&\langle \alpha _{\it N}\beta _{\it N}|
\tilde{v}_{\it N_{\rm 1}N_{\rm 2}}|
\gamma _{\it N}\delta _{\it N}\rangle
\end{eqnarray}
and
\begin{eqnarray}
\label{eq:def_S}
S^{({\it \Lambda N})}_{\alpha _{\it \Lambda}\beta _{\it N}
\gamma _{\it \Lambda}\delta _{\it N}}
&=&\langle \alpha _{\it \Lambda}\beta _{\it N}|
S_{\it \Lambda N_{\rm 1}}|\gamma _{\it \Lambda }
\delta _{\it N}\rangle ,\nonumber \\
S^{({\it NN})}_{\alpha _{\it N}\beta _{\it N}
\gamma _{\it N}\delta _{\it N}}
&=&\langle \alpha _{\it N}\beta _{\it N}|S_{\it N_{\rm 1}N_{\rm 2}}|
\gamma _{\it N}\delta _{\it N}\rangle .
\end{eqnarray}

The contributions of the TBC terms to
$\Delta E_{\lambda _{\it \Lambda}}^{(\rm TBC)}$
come from the effective three-body interactions
${\tilde{v}_{ijk}^{(v)}}$ and ${\tilde{v}_{ijk}^{(h)}}$.
We here note that the first and second terms of
${\tilde{v}_{ijk}^{(v)}}$ in Eq.~(\ref{eq:28}) do not bring about
non-zero contributions to the ${\it \Lambda}$ single-particle potential.
On the other hand, the contributions of ${\tilde{v}_{ijk}^{(h)}}$ in 
Eq.~(\ref{eq:29}) vanish up to  second order in $S_{ij}$. Therefore,
the contributions of ${\tilde{v}_{ijk}^{(v)}}$ and
${\tilde{v}_{ijk}^{(h)}}$ start with the second and third order terms
in $S_{ij}$, respectively.
These facts concerning the TBC terms can be proved by using
the property of decoupling of ${\tilde{v}_{ij}}$ and the restrictive
conditions for $S_{ij}$ in Eq.~(\ref{eq:restriction}) as discussed in
Ref.~\citen{rf:SO86}.
On the assumption that the matrix elements
of $S_{ij}$ are sufficiently small, we neglect the contribution of
${\tilde{v}_{ijk}^{(h)}}$.
The validity of this assumption will be confirmed numerically in \S 3.

The ${\it \Lambda}$ single-particle energy is given finally by
\begin{equation}
 E_{\lambda_{\Lambda}}=\langle \lambda_\Lambda |t_{\Lambda}|
\lambda_\Lambda\rangle
 +\langle \lambda_\Lambda |u_{\Lambda}^{(2)}| \lambda_\Lambda\rangle
 +\Delta E_{\lambda_{\Lambda}}^{({\rm p})}+\Delta
E_{\lambda_{\Lambda}}^{({\rm TBC})},
\end{equation}
where the first term is the kinetic energy, the second the
first-order potential energy, the third the sum of the perturbation
corrections given in Fig.~3 and the last the sum of the TBC contributions given
in Eq.~(\ref{eq:34}).

\subsection{Diagrammatical expression of the three-body-cluster term}

\begin{figure}[t]
  \epsfxsize = 8cm
  \centerline{\epsfbox{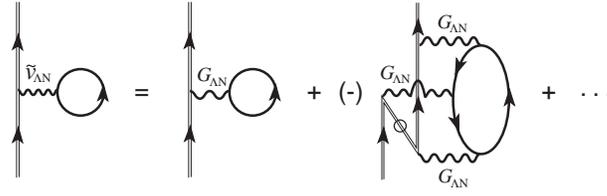}}
  \caption {Diagrammatical expression of the first-order ${\it \Lambda}$ potential energy
in terms of the ${\it \Lambda N}$ effective interaction $\tilde{v}_{\it \Lambda N}$
obtained in the UMOA and the $G$-matrix $G_{\it \Lambda N}$.}
  \label{fig:5}
\end{figure}

\begin{figure}[t]
  \epsfxsize = 8cm
  \centerline{\epsfbox{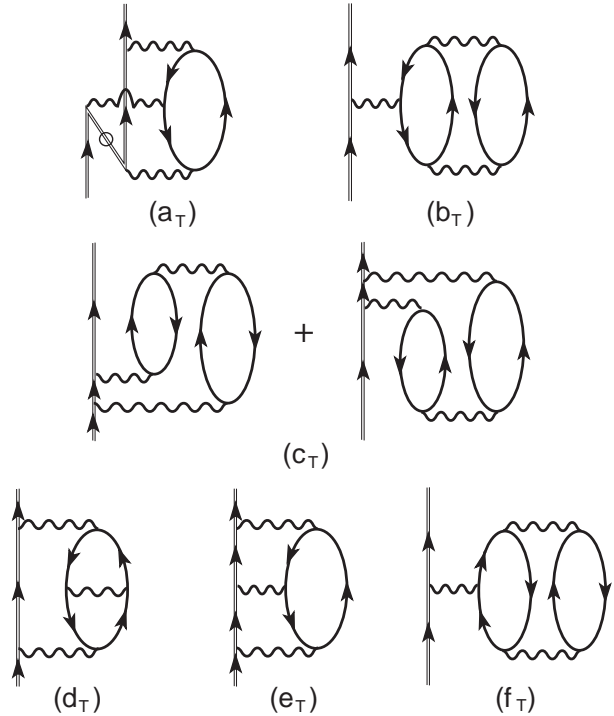}}
  \caption {Diagrams corresponding to the leading order terms of the
three-body-cluster terms in terms of the $G$-matrix shown by
the wavy line.}
  \label{fig:6}
\end{figure}

We here make some discussions on the relation between the UMOA and the
$G$-matrix theory in the calculation of the ${\it \Lambda}$
single-particle energy
in order to clarify the physical meaning of the TBC effect.
The formal relation between the effective interaction in the UMOA and
the $G$-matrix has been discussed in Refs.~\citen{rf:SO}
and \citen{rf:SO86}.
The ${\it \Lambda N}$ effective interaction $\tilde{v}_{\it \Lambda N}$
in the UMOA in the calculation of the first-order ${\it \Lambda}$
potential energy $u_{\it \Lambda}$ can be given in an expansion form
in terms of the $G$-matrix as shown diagrammatically in Fig.~4.
The second diagram on the right-hand side in Fig.~4 is a kind of
folded diagram.\cite{rf:KO} \ 
The leading contributions of the TBC terms in Eq.~(\ref{eq:35}) can
also be given diagrammatically in terms of the $G$-matrix as shown
in Fig.~5.
It is noted that the second diagram on the right-hand side in Fig.~4
has a minus sign which is against the diagram rule.
However, this diagram is just cancelled by diagram (a$_{\rm T}$)
in Fig.~5.
It should be noted that such a folded diagram emerges again as one of
higher-order corrections with the vertices of the
transformed interaction $ {\tilde v}_{ij}$.

\section{Numerical results for $_{\it \Lambda }^{17}$O and
$_{\it \Lambda }^{41}$Ca
using the Nijmegen and the J\"ulich ${\it YN}$ potentials}
\setcounter{equation}{0}

We performed calculations employing
the Nijmegen soft-core (NSC),\cite{rf:NSC} \ 
NSC97a-f,\cite{rf:NSC97} \ the J\"ulich \~A (J\~A) and
J\~B\cite{rf:JUL} potentials
for the ${\it YN}$ interaction.
We used the harmonic oscillator basis states
and took a common $ {\it \hbar \Omega } $ to single-particle states of
$ {\it \Lambda } $, $ {\it \Sigma } $ and $ {\it N} $.
We used the values $ {\it \hbar \Omega }=14 $MeV for
$_{\it \Lambda}^{17}$O and $12$MeV for $_{\it \Lambda}^{41}$Ca. 
The dependence of the calculated effective interaction and the
single-particle energy on the value
$ {\it \hbar \Omega } $ has already been examined,
and we confirmed that the dependence was quite small
around ${\it \hbar \Omega }=14 $MeV in $^{16}$O\cite{rf:SO86}
if perturbation corrections were included.
As for the nucleon single-particle potential $ u_{{\it N}} $,
we used the fixed
data\cite{rf:SO,rf:KSO} calculated using
the Paris potential,\cite{rf:Paris} \ 
and we have assumed that $ u_{{\it \Sigma }}=0 $.
The angular momentum $l$ of the partial wave of the ${\it NN}$ and
${\it YN}$ interactions is taken into account up to $l=4$.
The numbers $ \rho _{2}$,
$ \rho _{\rm F}$, $ \rho _{1} $ and $ \rho _{\rm X} $
are taken as
 $ \rho _{2}=1 $, $ \rho _{\rm F}=1 $, $ \rho _{1}=8 $
and $ \rho _{\rm X}=12 $ for $_{\it \Lambda }^{17}$O, and
$ \rho _{2}=2 $, $ \rho _{\rm F}=2 $, $ \rho _{1}=10 $
and $ \rho _{\rm X}=12 $ for $_{\it \Lambda }^{41}$Ca.

\begin{table}[b]
\caption{Typical matrix elements
$\langle aa|S_{\it NN}^{(2)}|bb \rangle _{J,T}$
and $\langle aa|S_{\it \Lambda N}^{(2)}|bb \rangle _{J,T=1/2}$ in
$_{\it \Lambda}^{17}$O for the two-body ${\it NN}$ and
${\it \Lambda N}$ correlation operators.
The Paris and  the NSC97f potentials are used as the ${\it NN}$ and
the ${\it YN}$ interactions, respectively.
The letters $a$ and $b$ denote single-particle orbits.
The $J$ and $T$ mean the total angular momentum and isospin,
respectively, for two-body states.}
  \label{table:6}
  \begin{center}
    \begin{tabular}{cccccrccr} \hline \hline
 $a$&$b$&$J$& &$T$&$\langle aa|S_{\it NN}^{(2)}|bb \rangle _{J,T}$
& &$T$&$\langle aa|S_{\it \Lambda N}^{(2)}|bb \rangle _{J,T}$\\
\cline{1-3} \cline{5-6} \cline{8-9}
  $1s_{1/2}$ & $0s_{1/2}$ & 1 & & 0 & $ 2.32\times 10^{-2}$ & &1/2
& $5.38\times 10^{-3}$\\
  $1s_{1/2}$ & $0s_{1/2}$ & 0 & & 1 & $ 1.69\times 10^{-2}$ & &1/2
& $9.42\times 10^{-3}$\\
  $1s_{1/2}$ & $0p_{1/2}$ & 1 & & 0 & $-2.43\times 10^{-2}$ & &1/2
& $2.49\times 10^{-3}$\\
  $1s_{1/2}$ & $0p_{1/2}$ & 0 & & 1 & $-5.75\times 10^{-3}$ & &1/2
& $9.36\times 10^{-4}$\\
  $0d_{3/2}$ & $0s_{1/2}$ & 1 & & 0 & $ 8.31\times 10^{-3}$ & &1/2
& $1.51\times 10^{-4}$\\
  $0d_{3/2}$ & $0s_{1/2}$ & 0 & & 1 & $ 7.82\times 10^{-3}$ & &1/2
& $4.58\times 10^{-3}$\\
  $0d_{3/2}$ & $0p_{1/2}$ & 1 & & 0 & $-6.48\times 10^{-3}$ & &1/2
& $3.67\times 10^{-3}$\\
  $0d_{3/2}$ & $0p_{1/2}$ & 0 & & 1 & $-8.85\times 10^{-3}$ & &1/2
& $1.29\times 10^{-3}$\\ \hline
    \end{tabular}
  \end{center}
\end{table}

The formulae for calculating the TBC contributions to
the ${\it \Lambda}$ single-particle energy have been derived in
Eq.~(\ref{eq:35}).
The TBC terms are given in powers of the two-body correlation operator
$S_{ij}$ as given in Eqs.~(\ref{eq:28}) and ~(\ref{eq:29}).
In the UMOA, it is necessary that the matrix element of
the correlation operator is sufficiently small in the sense of cluster
expansion.
In Table~\ref{table:6} typical matrix elements of $S_{ij}$ for
the ${\it NN}$ and ${\it \Lambda N}$ systems are given.
We present only the matrix elements of $S_{ij}$ of comparatively
large magnitudes among all the matrix elements of $S_{ij}$.
The magnitudes of the matrix elements
are at most of the order of $10^{-2}$.
Besides, the TBC effect yields the contributions of the order of
the square of $S_{ij}$ as shown in Eq.~(\ref{eq:35}).
Thus, we expect that the TBC contribution is considerably small
compared with the contribution from the one- and
two-body-cluster terms.

\begin{table}[b]
\caption{Contributions to the ${\it \Lambda}$ single-particle energies
in $_{\it \Lambda}^{17}$O for the NSC97f potential.
KE and PE stand for the kinetic and first-order potential
energies, respectively.
Rows (a), (b) and (c) are the contributions of diagrams
(a), (b) and (c) in Fig.~3,
respectively.
Rows (a$_{\rm T}$)-(f$_{\rm T}$) mean the contributions given in
Eq.~(\ref{eq:35}).
All entries are in MeV.}
\vspace{0.2cm}
\label{table:7}
  \begin{center}
    \footnotesize
    \begin{tabular}{crrrr} \hline \hline
        &0$s_{1/2} $ & 0$p_{3/2}$ & 0$p_{1/2}$ & $\Delta
\varepsilon_{ls}$\\ \hline
      KE        &  10.50 &  17.50 &  17.50 & *** \\
      PE        & -23.13 & -17.20 & -16.09 &  1.11 \\
     (a)        &  -1.84 &  -2.35 &  -2.42 & -0.08 \\
     (b)        &  -1.20 &  -0.37 &  -0.13 &  0.24 \\
     (c)        &  -0.55 &  -0.98 &  -1.25 & -0.28 \\
Subtotal        & -16.22 &  -3.40 &  -2.40 &  1.00\\ \hline
(a$_{\rm T}$)   &  -0.01 &  -0.01 &  -0.01 & *** \\
(b$_{\rm T}$)   &   1.49 &   1.22 &   1.15 & -0.07 \\
(c$_{\rm T}$)   &  -0.09 &  -0.16 &  -0.16 & *** \\
(d$_{\rm T}$)   &    ***   &   ***    &  -0.02 & -0.02 \\
(e$_{\rm T}$)   &   0.01 &   0.02 &   0.02 & *** \\
(f$_{\rm T}$)   &  -0.43 &  -0.46 &  -0.44 &  0.02\\
 Sum of TBC     &   0.97 &   0.60 &   0.54 & -0.06 \\ \hline
Total           & -15.25 &  -2.80 &  -1.86 & 0.95 \\ \hline
    \end{tabular}
  \end{center}
\end{table}

We calculated the ${\it \Lambda}$ single-particle energies
with the correction terms (a), (b) and (c) given in Fig.~3 and
the TBC terms (a$_{\rm T}$)-(f$_{\rm T}$) in Eq.~(\ref{eq:35}) for
$_{\it \Lambda}^{17}$O using the NSC97f potential as shown
in Table~\ref{table:7}.\footnote{
The mark *** in Tables~\ref{table:7} and \ref{table:8} means that
the calculated absolute value is less than 0.01MeV.}
One see that the TBC contributions are considerably small
in comparison with the contributions from the one- and two-body-cluster
terms.
This ensures that the cluster expansion method in powers of
the correlation operator $S_{ij}$ may be justified in the calculation of
the ${\it \Lambda}$ single-particle energy.

One also see that two of the TBC terms, (b$_{\rm T}$) and
(f$_{\rm T}$), have dominant contributions among the six TBC terms.
It may be clear from Fig.~5 that these two TBC terms
include the vertices inducing typical two-particle two-hole
excitations of the core nucleus in intermediate states.
On the other hand, the TBC terms (a$_{\rm T}$) and
(c$_{\rm T}$)-(e$_{\rm T}$) do not include this type of excitations.
The two terms (b$_{\rm T}$) and (f$_{\rm T}$), however,
give the contributions of the opposite sign.
Therefore, the sum of the TBC terms becomes rather small.
The effect of the TBC terms is, on the whole, repulsive and
at most about 3-4\% of the ${\it \Lambda}$ one-body potential
energies of the one- and two-body-cluster terms
for the 0$p$ and 0$s$ states in $_{\it \Lambda}^{17}$O as given
in Table~\ref{table:7}.

The calculated value of the spin-orbit splitting of the ${\it \Lambda}$
single-particle energies for the 0$p$ orbits in $_{\it \Lambda}^{17}$O
is $1.11$MeV for the first-order potential energy.
Corrections (a), (b) and (c) in Fig.~3 yield the contribution of
$-0.11$MeV to the ${\it \Lambda}$ spin-orbit splitting.
The sum of the TBC terms for the ${\it \Lambda}$ spin-orbit splitting
is only $-0.06$MeV.
We may conclude from the present calculation that the magnitude of
the ${\it \Lambda}$ spin-orbit splitting is determined dominantly by
the contribution of the one- and two-body-cluster terms, and
the polarization of the core nucleus does not play an important role.
This trend is also confirmed for the other ${\it YN}$ interactions used
in this study.

\begin{table}[b]
\caption{Results of $_{\it \Lambda}^{41}$Ca for the NSC97f.
The notation is the same as in Table~\ref{table:7}.
All entries are in MeV.}
\vspace{0.2cm}
\label{table:8}
  \begin{center}
    \footnotesize
    \begin{tabular}{crrrrrr} \hline \hline
        &0$s_{1/2} $ & 0$p_{3/2}$ & 0$p_{1/2}$ &
0$d_{5/2}$ & 0$d_{3/2}$ & 1$s_{1/2}$ \\ \hline
           KE   &   9.00 &  15.00 &  15.00 &  21.00 &  21.00 &  21.00\\
           PE   & -28.01 & -22.75 & -21.76 & -18.08 & -16.63 & -16.36\\
          (a)   &  -2.16 &  -3.19 &  -3.25 &  -3.69 &  -3.83 &  -2.96\\
          (b)   &  -1.44 &  -0.99 &  -0.87 &  -0.39 &  -0.15 &   0.16\\
          (c)   &  -0.58 &  -1.02 &  -1.25 &  -1.32 &  -1.73 &  -1.62\\
Subtotal        & -23.19 & -12.96 & -12.14 &  -2.48 &  -1.33 &   0.22\\
\hline
(a$_{\rm T}$)   & *** &  ***  &  ***   &   ***   &   ***   &    -0.01\\
(b$_{\rm T}$)   &   1.46 &   1.30 &   1.24 &   1.12 &   1.04 &   1.03\\
(c$_{\rm T}$)   &  -0.05 &  -0.08 &  -0.08 &  -0.13 &  -0.12 &  -0.15\\
(d$_{\rm T}$)   & *** &  ***  &   ***  &   ***   &   0.02  &     0.01\\
(e$_{\rm T}$)   &   0.01 &   0.01 &   0.01 &   0.02 &   0.02 &   0.02\\
(f$_{\rm T}$)   &  -0.38 &  -0.44 &  -0.44 &  -0.48 &  -0.46 &  -0.45\\
 Sum of TBC     &   1.03 &   0.78 &   0.74 &   0.53 &   0.49 &   0.46\\
\hline
Total           & -22.16 & -12.18 & -11.39 &  -1.95 &  -0.84 &   0.68\\
\hline
    \end{tabular}
  \end{center}
\end{table}

The calculated result in $_{\it \Lambda}^{41}$Ca for the NSC97f is
shown in Table~\ref{table:8}.
The effects of the second-order and the TBC correction terms
on the ${\it \Lambda}$ single-particle energy
are not so different from those in $_{\it \Lambda}^{17}$O.
The magnitude of the TBC terms is at most 3\%
of the ${\it \Lambda}$ one-body potential energy of the
one- and two-body-cluster terms for the 0$s$ state in
$_{\it \Lambda}^{41}$Ca.
We note here that the ${\it \Lambda}$ spin-orbit splittings are
calculated for both the
0$p$ and 0$d$ states in $_{\it \Lambda}^{41}$Ca.
The values of the ${\it \Lambda}$ spin-orbit splittings for the 0$p$
and 0$d$ states are $0.79$MeV and $1.11$MeV, respectively.
The net contribution of core polarizations to
the ${\it \Lambda}$ spin-orbit splittings is very small for both
the 0$p$ and 0$d$ states.
This situation is quite different from that in the ordinary nucleus
$^{40}$Ca,\cite{rf:KSO} \ 
as well as in $^{16}$O.\cite{rf:SO} \ 
It has been observed that many-body correlations bring about
corrections of more than $30$\% for the spin-orbit splittings of
nucleon states.

\begin{table}[b]
\caption{Partial-wave contributions to the first-order ${\it \Lambda}$
potential energy
for the 0$s_{1/2}$ state in $_{\it \Lambda}^{17}$O.
All entries are in MeV.}
  \label{table:9}
  \begin{center}
    \begin{tabular}{ccccccccc} \hline \hline
& & \hspace{-12mm}$l$ : even & & & & \hspace{-11mm}$l$ : odd & \\
\cline{2-3}\cline{5-8}
     & $^{1}S_{0}$ & $^{3}S_{1}$ & & $^{3}P_{0}$ & $^{1}P_{1}$
& $^{3}P_{1}$ & $^{3}P_{2}$ & Total \\ \hline
          NSC97a   &  -2.37 & -22.93 & & -0.03 &   0.84 &  0.80
& -0.98  & -24.68 \\
          NSC97b   &  -3.52 & -22.35 & &  0.04 &   0.90 &  0.98
& -0.88  & -24.82 \\
          NSC97c   &  -5.06 & -22.00 & &  0.18 &   0.95 &  1.12
& -0.78  & -25.60 \\
          NSC97d   &  -7.16 & -20.55 & &  0.30 &   1.06 &  1.40
& -0.56  & -25.52 \\
          NSC97e   &  -8.42 & -19.28 & &  0.35 &   1.14 &  1.62
& -0.41  & -24.99 \\
          NSC97f   &  -9.46 & -17.11 & &  0.32 &   1.29 &  2.00
& -0.14  & -23.10 \\ \hline
    \end{tabular}
  \end{center}
\end{table}

It might be of interest to compare the results obtained in this study
with those in other many-body methods.
Recently, the ${\it \Lambda}$ spin-orbit splittings in
$_{\it \Lambda}^{17}$O were calculated on another method by
Motoba\cite{rf:Mot} in which configuration mixing of
the ${\it \Lambda}$ single-particle states is
taken into account, using the YNG interaction \cite{rf:YB85}
derived from the NSC97 potentials.
The result for the NSC97f was almost $1$MeV for the 0$p$ states.
This value is consistent with our result of $0.95$MeV for the NSC97f
as given in Table~\ref{table:7}.

\begin{table}[b]
\caption{Partial-wave contributions to the first-order ${\it \Lambda}$
potential energy for the 0$s_{1/2}$ state in $_{\it \Lambda}^{41}$Ca.
All entries are in MeV.}
  \label{table:10}
  \begin{center}
    \begin{tabular}{ccccccc} \hline \hline
& & & & \hspace{-12mm}$l$ : even & & \\
\cline{2-7}
     & $^{1}S_{0}$ & $^{3}S_{1}$ & $^{3}D_{1}$ & $^{1}D_{2}$
& $^{3}D_{2}$ & $^{3}D_{3}$ \\ \hline
   J$\tilde{\rm A}$&  -3.55 & -23.76 & -0.12 &  -0.13 & -0.14 & -0.15\\
   J$\tilde{\rm B}$&   0.10 & -31.71 & -0.09 &  -0.13 & -0.14 & -0.11\\
          NSC97a   &  -2.80 & -29.42 & -0.09 &  -0.17 & -0.22 & -0.21\\
          NSC97b   &  -4.40 & -28.69 & -0.08 &  -0.19 & -0.23 & -0.22\\
          NSC97c   &  -6.57 & -28.25 & -0.08 &  -0.20 & -0.23 & -0.23\\
          NSC97d   &  -9.55 & -26.27 & -0.08 &  -0.23 & -0.24 & -0.24\\
          NSC97e   & -11.33 & -24.52 & -0.07 &  -0.24 & -0.25 & -0.24\\
          NSC97f   & -12.82 & -21.55 & -0.05 &  -0.26 & -0.27 & -0.24\\
          NSC      & -13.83 &  -8.92 & -0.06 &  -0.25 & -0.23 & -0.19\\
          NSC(*)   & -13.12 &  -9.50 & -0.05 &  -0.24 & -0.23 & -0.19\\
\hline
& & & \hspace{-12mm}$l$ : odd & & & \\
\cline{2-5}
 & $^{3}P_{0}$ & $^{1}P_{1}$ & $^{3}P_{1}$ & $^{3}P_{2}$ & & Total \\
\hline
    J$\tilde{\rm A}$  & 0.76 &  0.82 &  1.41 & -0.31 & & -25.17  \\
    J$\tilde{\rm B}$  & 0.70 &  1.16 &  2.00 &  0.52 & & -27.71  \\
          NSC97a      & -0.05 &  1.70 & 1.75 & -2.02 & & -31.53  \\
          NSC97b      &  0.09 &  1.83 & 2.14 & -1.83 & & -31.57  \\
          NSC97c      &  0.38 &  1.93 & 2.42 & -1.63 & & -32.48  \\
          NSC97d      &  0.63 &  2.16 & 3.02 & -1.21 & & -32.01  \\
          NSC97e      &  0.72 &  2.34 & 3.47 & -0.90 & & -31.03  \\
          NSC97f      &  0.67 &  2.67 & 4.26 & -0.36 & & -27.97  \\
          NSC         &  0.49 &  1.79 & 2.19 & -3.67 & & -22.69  \\
          NSC(*)      &  0.51 &  1.87 & 2.25 & -3.71 & & -22.42  \\
\hline
    \end{tabular}
  \end{center}
\end{table}

We now discuss the results obtained for various ${\it YN}$
interactions.
The partial-wave contributions of the ${\it \Lambda N}$ effective
interactions to the first-order ${\it \Lambda}$ potential energy
are given for various ${\it YN}$ potentials, namely,
the J\~A, J\~B, the NSC\footnote{The mark (*) for the NSC in
Table~\ref{table:10} and Fig.~6 means that the antisymmetric
spin-orbit force is not included.
This potential was also used in the calculation
of properties in $_{\it \Lambda}^{17}$O
in the previous work.\cite{rf:FOS99} \ 
On the other hand, the NSC newly used in the present study
is the complete version of the NSC potential code which includes the
antisymmetric spin-orbit force.}
and NSC97a-f potentials for the 0$s_{1/2}$ state,
in Table~\ref{table:9} for $_{\it \Lambda}^{17}$O, and
in Table~\ref{table:10} for $_{\it \Lambda}^{41}$Ca.
It is seen that the partial-wave contributions are very
different, especially in the spin singlet and triplet states,
dependently on the ${\it YN}$ potentials employed.
It is noted, however, that the ratios of the contributions of
the spin singlet and triplet states vary smoothly from the NSC97a
to the NSC97f.
This variation is caused mainly by the difference
in the strength of the spin-spin interaction of the NSC97 potentials.
The similar tendency was observed in the calculation of the
potential energy of ${\it \Lambda}$ contained
in nuclear matter.\cite{rf:NSC97} \ 

One of the features of the NSC97 model for the ${\it YN}$ interaction
is that the NSC97a-f potentials are given by varying smoothly
the magnetic $F/(F+D)$ ratio $\alpha _{V}^{m}$ for the vector mesons
in the NSC97a-f potentials.
This change in the magnetic ratio yields the variation of the strength
of not only the spin-spin interaction but also other components
including the spin-orbit interaction.
According to the variation of the ${\it YN}$ interactions,
the calculated result of the ${\it \Lambda}$ single-particle potential
energies vary, dependently on the ${\it YN}$ potentials employed
as seen in Tables~\ref{table:9} and \ref{table:10}.
It is important to investigate the difference among the NSC97
potentials quantitatively in the calculation of properties in finite
${\it \Lambda}$ hypernuclei.
This quantitative study would help to improve the ${\it YN}$
interaction, complementarily to the progress in relevant experiments of
${\it \Lambda}$ hypernuclei.

\begin{figure}[t]
  \epsfxsize = 10cm
  \centerline{\epsfbox{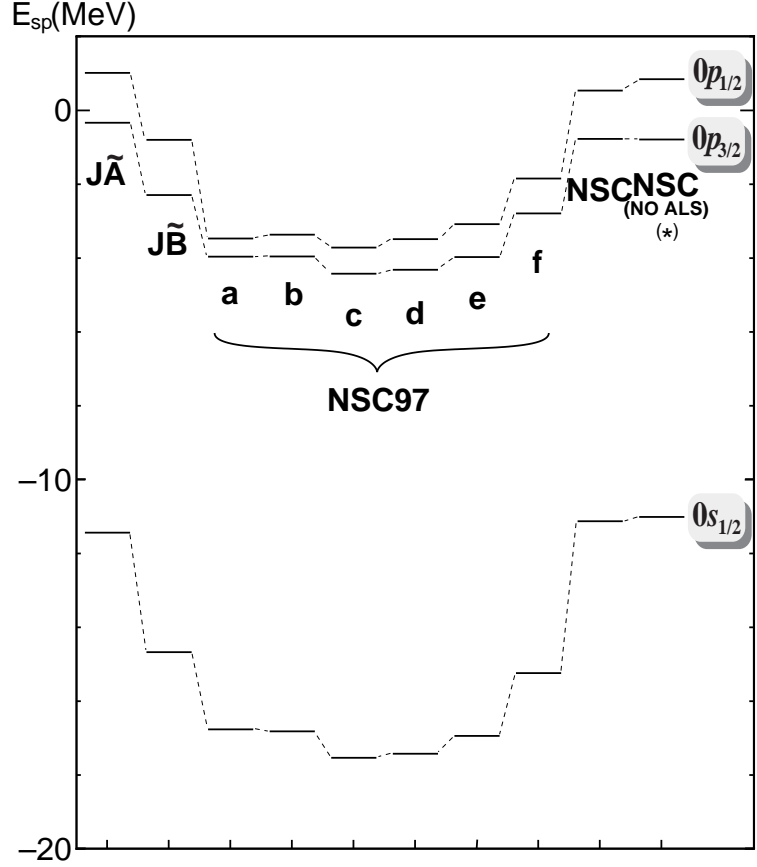}}
  \caption{Dependence of the calculated ${\it \Lambda}$ single-particle
energies in $_{\it \Lambda}^{17}$O
on various ${\it YN}$ interactions. All entries are in MeV.}
\label{fig:8}
\end{figure}

\begin{figure}[t]
  \epsfxsize = 10cm
  \centerline{\epsfbox{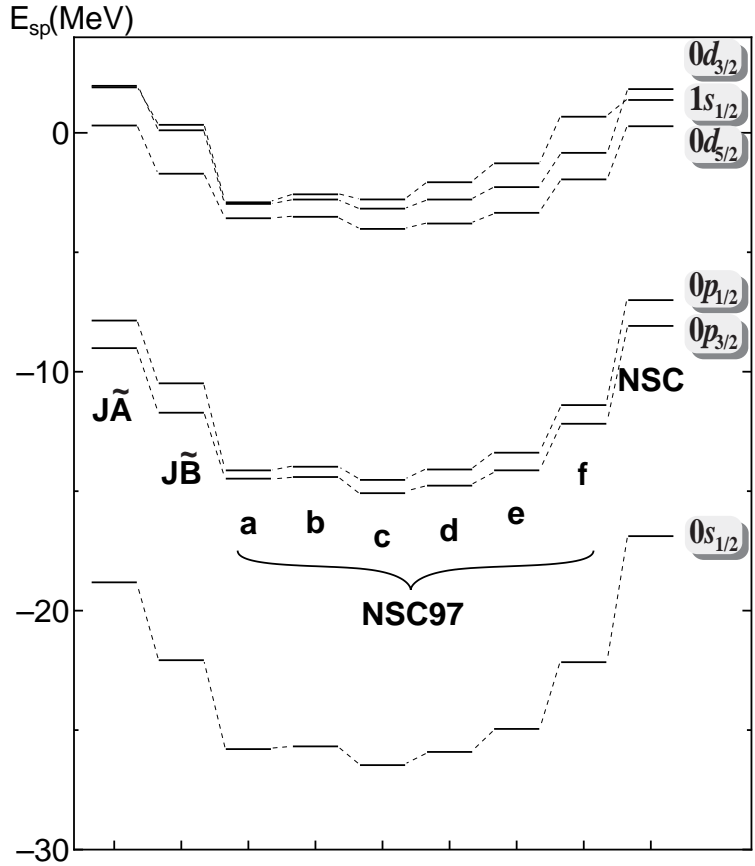}}
  \caption{Dependence of the calculated ${\it \Lambda}$ single-particle
energies in $_{\it \Lambda}^{41}$Ca
on various ${\it YN}$ interactions. All entries are in MeV.}
\label{fig:9}
\end{figure}

In Figs.~6 and 7 we show the calculated ${\it \Lambda}$ single-particle
energies including the second-order
perturbation corrections and the TBC terms for the J\~A, J\~B, the NSC
and  NSC97 interactions.
It is noted here that the calculated results for the NSC97f potential
are, on the whole, fairly good in comparison with the empirical
values for $_{\it \Lambda}^{16}$O and
$_{\it \Lambda}^{40}$Ca.\cite{rf:VID98} \ 
Our result for the NSC97f is consistent with the fact that
the calculations of few-body ${\it \Lambda}$ hypernuclei
such as $_{\it \Lambda}^{3}$H\cite{rf:NSC97} and
$_{\it \Lambda}^{4}$He\cite{rf:Akaishi}
using the NSC97f can reproduce successfully experimental energy levels.

The calculated ${\it \Lambda}$ spin-orbit splittings of the 0$p$ states
vary in the range of  $0.49$-$1.31$MeV from the NSC97a to the NSC in
$_{\it \Lambda}^{17}$O as shown in Fig.~6.
Similarly, the ${\it \Lambda}$ spin-orbit splittings of the 0$p$ and
0$d$ states vary in the range of $0.35$-$1.08$MeV and $0.61$-$1.55$MeV,
respectively, in $_{\it \Lambda}^{41}$Ca as shown in Fig.~7.
The experimental values of the spin-orbit splitting of a nucleon are
known as 6.20MeV for the 0$p$ states in $^{16}$O and 6.75MeV for the
0$d$ states in $^{40}$Ca.
Therefore, the obtained values of the ${\it \Lambda}$ spin-orbit
splitting are considerably smaller than those of a nucleon.
Recently, some experiments in connection with the ${\it \Lambda}$
spin-orbit splittings
have been performed at KEK for $_{\it \Lambda}^{89}$Y and
$_{\it \Lambda}^{51}$V\cite{rf:Nag95}
and at BNL for $_{\it \Lambda}^{13}$C\cite{rf:Kis96} and
$_{\it \Lambda}^{9}$Be.\cite{rf:Tam96} \
Higher resolution
has been attained in these experiments than that in old
experiments though the analysis of these new experiments
are still being in progress. Therefore, it will be highly desirable
that the magnitude of the ${\it \Lambda}$ spin-orbit splitting is
established experimentally.

\section{Concluding remarks}

A method of evaluating the three-body-cluster (TBC) effect
was presented for the calculation of
${\it \Lambda}$ single-particle energies.
The TBC terms were generated as the transformed three-body interactions
among the ${\it YNN}$ and ${\it NNN}$ systems.
The transformed three-body interactions were given in the cluster
expansion form
in powers of two-body correlation operators of the ${\it YN}$ and
${\it NN}$ systems.
Therefore, it is necessary for fast convergence of the cluster
expansion that the matrix elements of the correlation
operator should be sufficiently small.
In our calculations, it was made sure that the matrix elements of
the correlation operator were quite small, and thus the TBC
contributions to the ${\it \Lambda}$ single-particle energies were
considerably small in comparison with the contributions from
the one- and two-body-cluster terms.
This ensured that the cluster expansion method in terms of the
correlation operator was acceptable in the calculation of
${\it \Lambda}$ single-particle energies.

The present results showed that $ {\it \Lambda } $ in the 0$p_{1/2}$
state in $_{\it \Lambda }^{17}$O was not bound for the J\~A and the NSC
potentials but bound for the J\~B and the NSC97a-f potentials.
The spin-orbit splittings of the ${\it \Lambda}$ single-particle orbits
in the 0$p$ states,
including the TBC contributions,
were given  by $1.35$,
$1.49$, $1.31$ and $0.49$-$0.95$MeV for the J\~A, J\~B, the NSC and
NSC97a-f potentials, respectively.
The value of 1.49MeV for the J\~B potential is very large compared with
that given in Ref.~\citen{rf:VID98}.
On the other hand, the value 0.95MeV for the NSC97f potential
is consistent with that given in Ref.~\citen{rf:Mot}.
The calculated ${\it \Lambda}$ spin-orbit splittings depend largely on
the ${\it YN}$ interactions employed.
The dependence of the calculated results on the ${\it YN}$ interactions
was also examined for $_{\it \Lambda }^{41}$Ca.
It seemed that the NSC97f potential yielded better results for
$_{\it \Lambda }^{17}$O and $_{\it \Lambda }^{41}$Ca
in comparison with the relevant experimental values.
The present results were consistent with those
obtained in the calculation of few-body ${\it \Lambda}$ hypernuclei.

In the application of the UMOA to ${\it \Lambda}$ hypernuclei,
some problems still remain.
First, in the present calculation, single-particle energies of
$ {\it \Sigma } $ in intermediate states were neglected.
However, the single-particle spectrum of $ {\it \Sigma } $ has a
possibility of bringing about some contributions to the
$ {\it \Lambda N} $ effective interaction.
Although single-particle energies of $ {\it \Sigma } $ has not yet
been known well,
it would be important to investigate how much the spectrum of
$ {\it \Sigma } $ affects the determination of the $ {\it \Lambda N} $
effective interaction.
Second, there would be some relativistic effect for the ${\it \Lambda}$
single-particle energy as well as in calculations of ordinary nuclei.
In the present approach, the relativistic effect has not yet
been considered
in studies of ordinary nuclei and hypernuclei.
On the other hand, the relativistic many-body approach such as
the Dirac-Brueckner-Hartree-Fock approach\cite{rf:rel} does not fully
take account of the effect of three-or-more-body correlations.
Therefore, it would be of great importance for the understanding of
nuclear properties from the many-body theoretical point of view
that the relativistic effect is incorporated in the framework of
the UMOA.
This problem, which would be closely related to the treatment
of the three-body force,\cite{rf:3fW,rf:3fP} \ 
is also a remaining task in the UMOA.

\section*{Acknowledgements}
One of the authors (S.~F.) would like to thank Prof. K.~Takada,
Prof. M. Kamimura and Prof.~Y.~R. Shimizu
for their continuous encouragement and instructive discussions at
Kyushu University.
We are grateful to Prof. T.~Motoba for stimulating and helpful
discussions on $_{\it \Lambda}^{17}$O which is one of the main
interests in this study.
We also thank Prof. Y.~Akaishi, Prof. Y.~Yamamoto, Prof. Th. A. Rijken
and Prof. K.~Miyagawa for useful discussions and their help for using
the Nijmegen and the J\"ulich ${\it YN}$ potential codes.


%

\end{document}